***Unit Cell Volume, and Lattice Parameter of Cubic High Entropy Alloys***

J. Coreño Alonso[a], O. Coreño Alonso[a]*.

[a] Universidad de Guanajuato, Juárez # 77 Col. Centro, C.P. 36,000, Guanajuato, Gto. México.

*ocoreno@ugto.mx.

**Abstract.**

An equation has been derived to predict unit cell volume of high entropy alloys, HEA´s, by two different methods. Both treatments led to the same equation. For cubic HEA´s lattice parameters were calculated. The predicted lattice parameters were compared with those reported for 68 HEA´s. Lattice parameters were also calculated using the equivalent of Vegard´s law for these alloys. Average errors were 0.52%, and 0.42% when Vegard´s law, and the equation derived in this work were used, respectively.

Preprint

**1.- Introduction.**

High entropy alloys, HEA´s, have received a lot of attention in recent years, and the possible use of high entropy alloys in diverse fields has been reported. Some of the possible applications are in catalysis [1], as ferromagnetic materials [2], in biomedical implants [3], in magnetocaloric refrigeration [4], as corrosion resistant materials [5], tribological materials [6], antiviral material [7], as an oxide dispersion-strengthened alloy [8], superconductor material [9], irradiation resistant materials [10], or as an alloy with transformation-induced plasticity [11].

Expressions to calculated lattice parameter in cubic binary or ternary intermetallics, with $L1_2$ [20] or B2 [17] structures, have been derived. From the equations in these references, an expression to calculate the unit cell volume of intermetallic compounds of any crystalline structure, and with any number of different solute element atoms was derived [21]. In this reference, it was proposed to take the element with highest concentration as the solvent. Data of intermetallic compounds with up to three elements were employed to evaluate the model. Good results were obtained with this approximation, even when solvent concentration was as low as 50% in some cases. If this approximation is to be applied to HEA´s, a problem arises due to the difficulty to define the solvent when none of the elements has concentration higher than 50%. In this work, two alternatives have been developed to tackle this problem. First, a HEA is considered as a mixture of solid solutions, later, a HEA is modeled as a mixture of solute element atoms surrounded by different environments. Both treatments led to the same equation to calculate unit cell volume. For cubic HEA´s, lattice parameter can then be calculated. The predicted lattice parameters were compared with those reported for 68 HEA´s.

## 2.- Equation to Calculate Unit Cell Volume for HEA´S Using Volume Size Factors.

Figure 1 shows a simplified scheme of a compact plane in a quaternary cubic HEA. It can be observed that atoms with the highest concentration, fifty percent, are blue. Red, black, and green atoms have half the ratio of blue atoms. In this figure, it is easy to identify the solvent element since it is the one with the highest concentration. Based on this figure, a HEA was considered as a mixture of solid solutions with different compositions. Then, an equation to calculate unit cell volume of the HEA was derived.

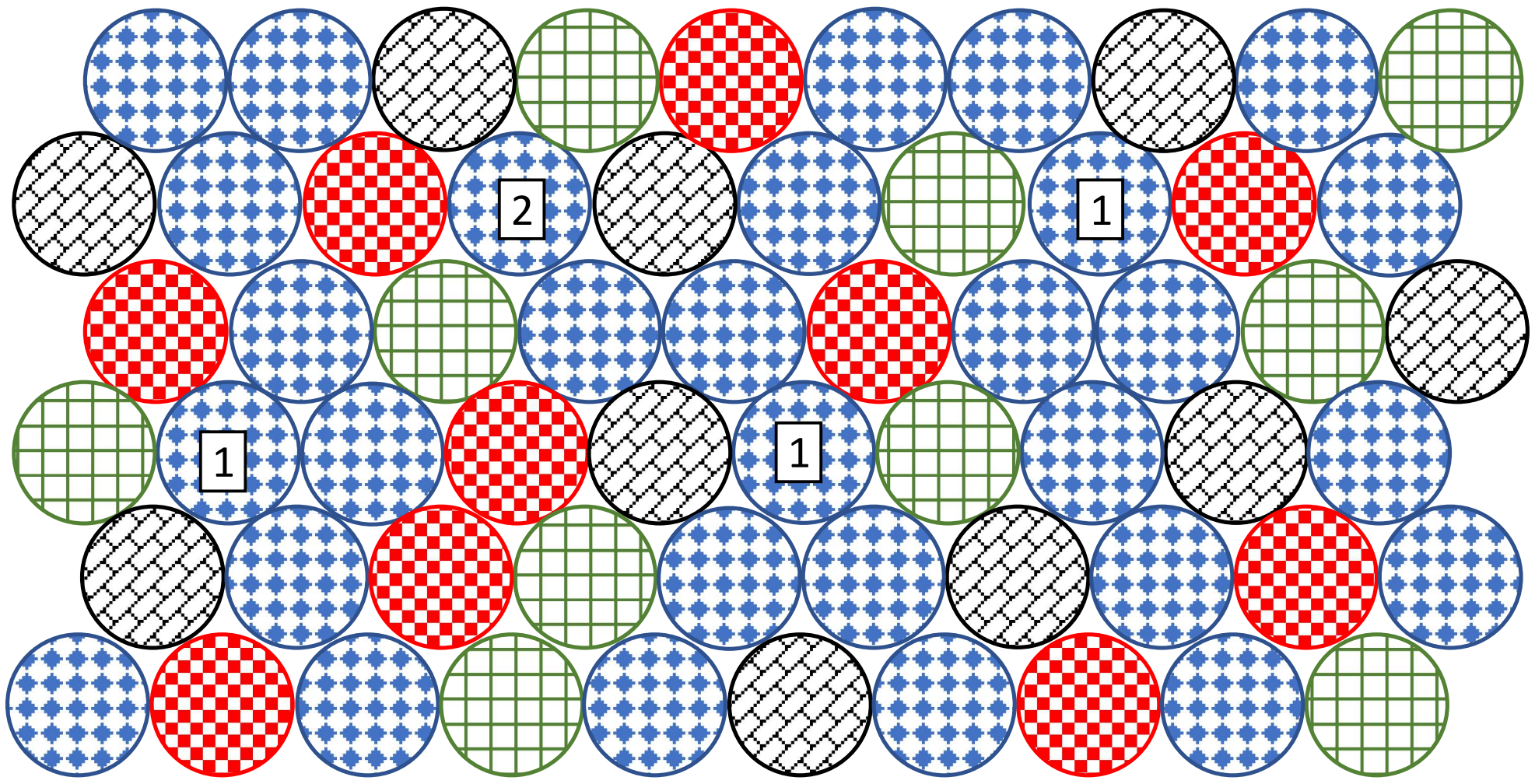

Figure 1. Simplified scheme of a compact plane in a quaternary cubic HEA.

The unit cell volume of a solid solution or intermetallic compound $A_XB_YC_{W..}M_l$ can be expressed as [21]

$$V_{A_xB_yC_W\ldots M_l} = Z\left(x\bar{V}_A + y\bar{V}_B + w\bar{V}_C + \cdots + l\bar{V}_M\right) \quad (1)$$

where $Z = \frac{Number\ of\ atoms\ in\ the\ unit\ cell}{(x+y+w+..+l)}$ (2)

If the element A is considered as the solvent element, then, from the difinition of volume size factor, $\Omega_{sf}$, [22] the effective atomic volume of element *i* in the solid solution, $\overline{V_i}$, can be expressed as

$$\overline{V_i} = V_A(1 + \Omega_{sfi/A}) \quad (3)$$

If (3) is replaced in (1) for each *i* element, then

$$V_{A_xB_yC_WM_l} = Z\,V_A\big[(x + y + w + \cdots + l) + (y\Omega_{sfB/A} + w\Omega_{sfC/A} + \cdots + l\Omega_{sfM/A})\big] = unit\ cellvolume\ taking\ A\ as\ solvent\ element \quad (4)$$

If B, C, or M are considered as the solvent element, then the volume of the unit cell of the solid solution can be expressed, respectively, as

$$V_{A_xB_yC_w\ldots M_l} = Z\,V_B\,\big[(x + y + w + \cdots + l) + (x\Omega_{sfA/B} + w\Omega_{sfC/B} + \cdots + l\Omega_{sfM/B})\big] = unit\ cell\ volume\ taking\ B\ as\ solvent\ element \quad (5)$$

$$V_{A_xB_yC_W\ldots M_l} = Z\,V_C\big[(x + y + w + \cdots + l) + (x\Omega_{sfA/C} + y\Omega_{sfB/C} + \cdots + l\Omega_{sfM/C})\big] = unit\ cell\ volume\ taking\ C\ as\ solvent\ element \quad (6)$$

•

•

$$V_{A_xB_yC_W\ldots M_l} = Z\,V_M\big[(x + y + w + \cdots + l) + (x\Omega_{sfA/M} + y\Omega_{sfB/M} + \mathrm{w}\Omega_{sfC/M+\ldots})\big] = unit\ cell\ volume\ taking\ M\ as\ solvent\ element \quad (7)$$

Take $c_i$ as the concentration of each element in the solid solution. If we consider the solid solution as a mixture of each solid solution (4) to (7), then the volume of the unit cell of this mixture is

$$V_{A_xB_yC_W\ldots M_l} = c_AV\ unit\ cell\ A\ as\ solvent + c_BV\ unit\ cell\ B\ as\ solvent + c_CV\ unit\ cell\ C\ as\ solvent + \cdots + c_MV\ unit\ cell\ M\ as\ solvent \quad (8)$$

$$V_{A_xB_yC_W\ldots M_l} = Zc_A\,V_A\big[(x + y + w + \cdots + l) + (y\Omega_{sfB/A} + w\Omega_{sfC/A} + \cdots + l\Omega_{sfM/A})\big]$$

$$+ Zc_B\, V_B\,\left[(x + y + w + \cdots + l) + (x\Omega_{sfA/B} + w\Omega_{sfC/B} + \cdots + l\Omega_{sfM/B})\right] + Zc_c\, V_C\left[(x + y + w + \cdots + l) + (x\Omega_{sfA/C} + y\Omega_{sfB/C} + \cdots + l\Omega_{sfM/C})\right] + \cdots + Zc_M\, V_M\left[(x + y + w + \cdots + l) + (x\Omega_{sfA/M} + y\Omega_{sfB/M} + w\Omega_{sfC/M+\cdots})\right] \quad (9)$$

$$V_{A_xB_yC_W\ldots M_l} = Number\ of\ atoms\ in\ the\ unit\ cell * [c_A\, V_A + c_B\, V_B\ +\ c_c\, V_C + \cdots + c_M\, V_M] + Zc_A\, V_A\left[y\Omega_{sfB/A} + w\Omega_{sfC/A} + \cdots + l\Omega_{sfM/A}\right] +\ Zc_B\, V_B\,\left[x\Omega_{sfA/B} + w\Omega_{sfC/B} + \cdots + l\Omega_{sfM/B}\right] + Zc_c\, V_C\left[x\Omega_{sfA/C} + y\Omega_{sfB/C} + \cdots + l\Omega_{sfM/C}\right] + \cdots + Zc_M\, V_M\left[x\Omega_{sfA/M} + y\Omega_{sfB/M} + \mathrm{w}\Omega_{sfC/M+\cdots}\right] \quad (10)$$

For equiatomic high entropy alloys, $c_i$ = $c$ for all elements, and (10) can be simplified to

$$V_{A_xB_yC_W\ldots M_l} = Number\ of\ atoms\ in\ the\ unit\ cell * c * [\, V_A +\ V_B\ +\ V_C + \cdots +\ V_M] + Zc * \left(\, V_A\left[y\Omega_{sfB/A} + w\Omega_{sfC/A} + \cdots + l\Omega_{sfM/A}\right] +\ V_B\,\left[x\Omega_{sfA/B} + w\Omega_{sfC/B} + \cdots + l\Omega_{sfM/B}\right] + V_C\left[x\Omega_{sfA/C} + y\Omega_{sfB/C} + \cdots + l\Omega_{sfM/C}\right] + \cdots + V_M\left[x\Omega_{sfA/M} + y\Omega_{sfB/M} + \mathrm{w}\Omega_{sfC/M+\cdots}\right]\right) \quad (11)$$

Alternatively, the unit cell volume of the high entropy alloy could be derived considering each solute atom in the alloy.

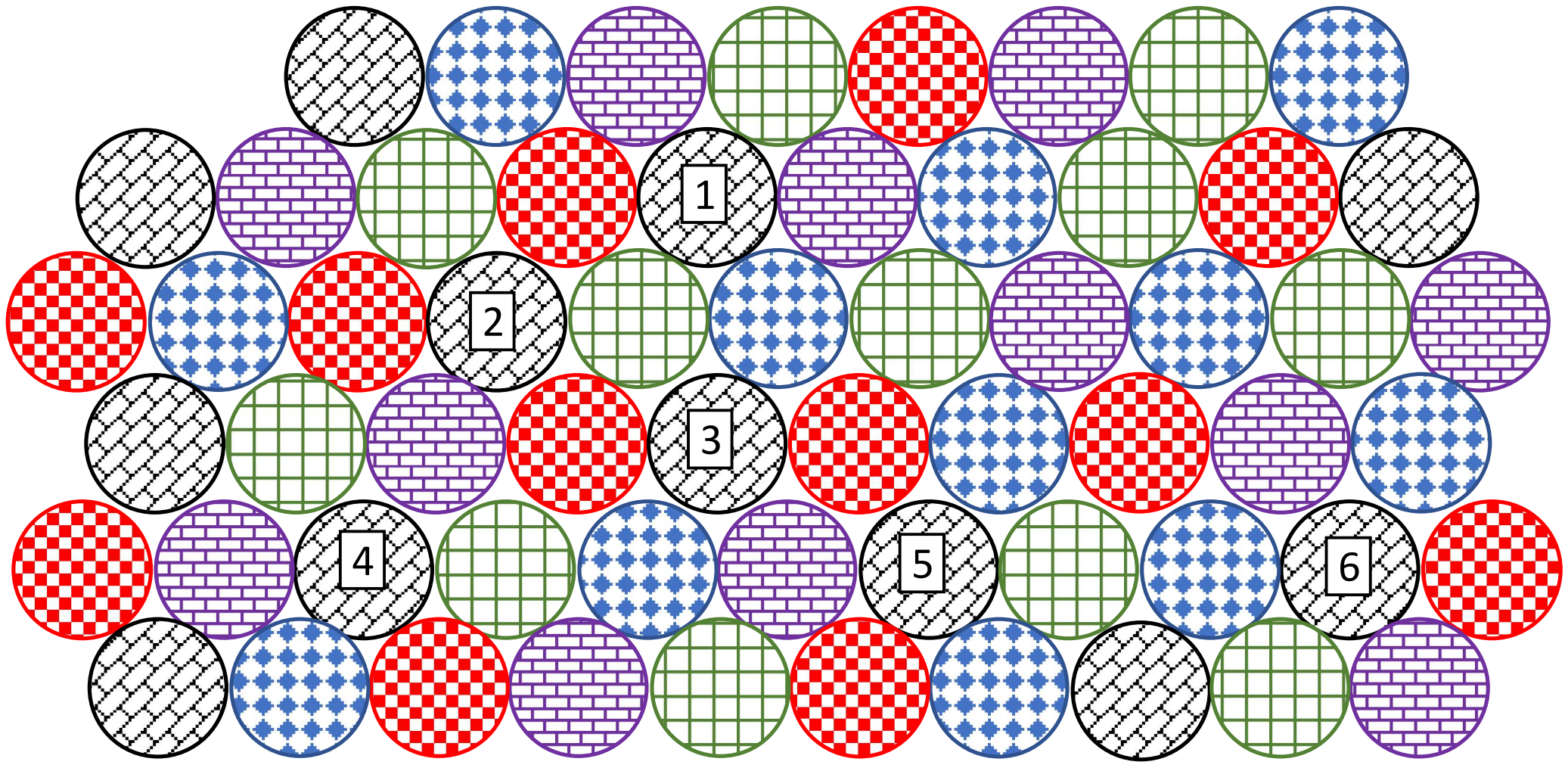

Figure 2. Simplified scheme of a compact plane in an equiatomic quinary cubic HEA.

In the solid solution $A_XB_YC_{W..}M_l$, the effective atomic volume, $\overline{V_\iota}$, of each solute element in the corresponding solvent is reported in Table 1.

Table 1. Effective atomic volume, $\overline{V_i}$, of each solvent element in the corresponding solute, in the solid solution $A_X B_Y C_{W..} M_l$.

| Solute \ Solvent | *A* | *B* | *C* | *M* |
|---|---|---|---|---|
| *A* | $V_A$ | $V_B(1+\Omega_{sfA/B})$ | $V_C(1+\Omega_{sfA/C})$ | $V_M(1+\Omega_{sfA/M})$ |
| *B* | $V_A(1+\Omega_{sfB/A})$ | $V_B$ | $V_C(1+\Omega_{sfB/C})$ | $V_M(1+\Omega_{sfB/M})$ |
| *C* | $V_A(1+\Omega_{sfC/A})$ | $V_B(1+\Omega_{sfC/B})$ | $V_C$ | $V_M(1+\Omega_{sfC/M})$ |
| *M* | $V_A(1+\Omega_{sfM/A})$ | $V_B(1+\Omega_{sfM/B})$ | $V_C(1+\Omega_{sfM/C})$ | $V_M$ |

Given that the solvent elements concentrations are $C_A$, $C_B$, $C_{D,,}$, and $C_M$, the average atomic volume of each solute element, $V_{i,av}$, is

$$V_{A,av} = C_A V_A + C_B V_B(1+\Omega_{sfA/B}) + C_C V_C(1+\Omega_{sfA/C}) + \cdots + C_M V_M(1+\Omega_{sfA/M}) \quad (12)$$

$$V_{B,av} = C_A V_A(1+\Omega_{sfB/A}) + C_B V_B + C_C V_C(1+\Omega_{sfB/C}) + \cdots + C_M V_M(1+\Omega_{sfB/M}) \quad (13)$$

$$V_{C,av} = C_A V_A(1+\Omega_{sfC/A}) + C_B V_B(1+\Omega_{sfC/B}) + C_C V_C + \cdots + C_M V_M(1+\Omega_{sfC/M}) \quad (14)$$

$$V_{M,av} = C_A V_A(1+\Omega_{sfM/A}) + C_B V_B(1+\Omega_{sfM/B}) + C_C V_C(1+\Omega_{sfM/C}) + \cdots \quad (15)$$

Equations (12) - (15) would describe the average volume, that a solute atom has when it is surrounded by different number of atoms, of other elements. The high entropy solid solution would be described as a mixture of solute atoms, in several different surrounding environments.

The volume of each element in the unit cell, on average, is by $C_i V_{i,av}$, and the volume of the unit cell can be expressed as:

$$V_{A_x B_y C_W \ldots M_l} = Number\ of\ atoms\ in\ the\ unit\ cell * \left[C_A\, V_{A,av} + C_B\, V_{B,av} + C_C\, V_{C,av} + \cdots + C_M\, V_{M,av}\right] \quad (16)$$

Let us represent the *number of atoms in the unit cell as nauc.* Then,

$$V_{A_x B_y C_W \ldots M_l} = nauc * [C_A V_A + C_B V_B + C_C V_C + \cdots + C_M V_M] * (C_A + C_B + C_C + \cdots + C_M) + nauc * C_A V_A\left(C_B \Omega_{sfB/A} + C_C \Omega_{sfC/A} + \cdots + C_M \Omega_{sfM/A}\right) + nauc * C_B V_B\left(C_A \Omega_{sfA/B} + C_B \Omega_{sfC/B} + \cdots + C_M \Omega_{sfM/B}\right) + nauc * C_C V_C\left(C_A\, \Omega_{sfA/C} + C_B \Omega_{sfB/C} + \cdots + C_M \Omega_{sfM/C}\right) + nauc * C_M V_M (C_A \Omega_{sfA/M} + C_B V_B \Omega_{sfB/M} + C_C \Omega_{sfC/M+\cdots}) \quad (17)$$

Given that $C_A = x/(x+y+w+\cdots+l)$, $C_B = y/(x+y+w+\cdots+l)$, $C_C = w/(x+y+w+\cdots+l)$, ..., $C_M = l/(x+y+w+\cdots+l)$ (18)

and that $C_A + C_B + C_C + \cdots + C_M = 1$ (19)

substituting (18) in (17), rearranging, and using (19), equation (10) is obtained again.

$$V_{A_xB_yC_W\ldots M_l} = Number\ of\ atoms\ in\ the\ unit\ cell * [c_A V_A + c_B V_B + c_c V_C + \cdots + c_M V_M] + Zc_A V_A[y\Omega_{sfB/A} + w\Omega_{sfC/A} + \cdots + l\Omega_{sfM/A}] + Zc_B V_B [x\Omega_{sfA/B} + w\Omega_{sfC/B} + \cdots + l\Omega_{sfM/B}] + Zc_c V_C[x\Omega_{sfA/C} + y\Omega_{sfB/C} + \cdots + l\Omega_{sfM/C}] + \cdots + Zc_M V_M[x\Omega_{sfA/M} + y\Omega_{sfB/M} + w\Omega_{sfC/M} \ldots] \quad (10)$$

As can be seen, the same expression is obtained by the two methods described above.

**3.- Discussion.**

To evaluate equation (10), calculated values were compared with reported values of lattice parameters of 68 cubic HEA´s. Unit cell volumes were converted to lattice parameters. Results are shown in Table 2. Volume size factors were taken from [22]. If they were not reported, they were calculated from crystallographic data. For eight systems crystallographic data were no found. Then, $\Omega_{sf}$ was interpolated from the least square linear equation of $\Omega s_f$ vs. solute atomic volume. This, for reported $\Omega_{sf}$ values of the solvent element. The corresponding values are reported in Table 3.

Table 2. Reported, and calculated lattice parameters of HEA´s, and the corresponding errors.

| High Entropy Alloy | $a_0$ reported | $a_0$ Vegard´s law | $a_0$ equation (10) | Vegard´s law error | equation (10) error | $a_0$ determined by | Chemical composition by |
|---|---|---|---|---|---|---|---|
| W.273Nb.227Mo.256Ta.244 [23] | 3.2134 | 3.2263 | 3.2148 | 0.401 | 0.044 | XRD | ICP -OES |
| W.25Nb.22Mo.26Ta.27 [24] | 3.24 | 3.23 | 3.22 | 0.309 | 0.617 | XRD | EDS |
| NbMoTaW [25] | 3.222 | 3.231 | 3.217 | 0.279 | 0.155 | XRD | MNR |
| W.211Nb.206Mo.217Ta.156V.210 [23] | 3.1832 | 3.1849 | 3.1804 | 0.053 | 0.088 | XRD | ICP -OES |
| WNbMoTaVTi [26] | 3.216 | 3.209 | 3.188 | 0.218 | 0.871 | XRD Bragg´s Law | EDS |
| CoFeReRu [27] v/z | 25.482 | 25.585 | 25.436 | 0.404 | 0.180 | XRD | MNR |
| CoCrFeNi [28] | 3.575 | 3.579 | 3.587 | 0.112 | 0.336 | XRD WPPM | MNR |
| CoCrFeNiMn [28] | 3.597 | 3.594 | 3.602 | 0.083 | 0.139 | XRD WPPM | MNR |
| Co.204Cr.205Fe.202Mn.194Ni.195 [29] | 3.59 | 3.59 | 3.6 | 0.000 | 0.279 | XRD | EPMA |
| Co.203Cr.194Fe.206Mn.201Ni.196 [30] | 3.60 | 3.59 | 3.60 | 0.278 | 0.000 | XRD | XRF |
| Cr.127Fe.498Ni.111Mn.264 [31] | 3.61 | 3.62 | 3.62 | 0.277 | 0.277 | XRD | EDS |
| Co.20Cr.20Fe.40Ni.10Mn.10 [32] | 3.587 | 3.598 | 3.605 | 0.307 | 0.502 | XRD | MNR |
| Co.211Cr.187Fe.342Ni.063Mn.197 [33] | 3.588 | 3.605 | 3.611 | 0.474 | 0.641 | XRD | MNR |
| Ru.185Rh.156Pd.182Os.143Ir.159Pt.174 [34] | 3.8473 | 3.8462 | 3.8471 | 0.029 | 0.005 | XRD Rietveld | XRF |
| Ru.19Rh.20Re.21Os.21Ir.19 V/Z [35] | 13.979 | 14.042 | 14.006 | 0.451 | 0.193 | XRD Rietveld | EDS |
| NbZrHfVTi [36] | 3.377 | 3.361 | 3.374 | 0.474 | 0.089 | XRD Rietveld | MNR |
| Al.262Cr.241Fe.259Mo.235V.003 [37] | 3.01 | 3.04 | 3.02 | 0.997 | 0.332 | XRD * | EDS |
| Al.200Cr.243Fe.232Mo.166V.158 [37] | 2.98 | 3.02 | 3.01 | 1.342 | 1.007 | XRD* | EDS |
| Al.180Ni.189Cu.214Fe.196Cr.220 [38] | 2.894 | 2.926 | 2.918 | 1.106 | 0.829 | XRD | EDS |
| CoCrCuNiZn [39] | 2.8831 | 2.9012 | 2.8815 | 0.628 | 0.056 | XRD | MNR |
| NbZrHfTi [40] | 3.438 | 3.435 | 3.428 | 0.087 | 0.291 | XRD | MNR |
| Nb.200Mo.208Cr.187Ti.202V.203 [41] | 3.140 | 3.139 | 3.138 | 0.032 | 0.064 | XRD | EDS |
| NbTaTiV [42] | 3.23 | 3.23 | 3.23 | 0.000 | 0.000 | XRD | MNR |
| NbTaTiV [43] | 3.2206 | 3.2319 | 3.2299 | 0.351 | 0.289 | Neutron diffraction, refined | MNR |
| Nb.27Mo.21Ta.27W.24 [44] | 3.218 | 3.226 | 3.213 | 0.249 | 0.155 | XRD Nelson-Riley regression | EDS |
| Nb.22Mo.18Ta.28W.31 [44] | 3.216 | 3.222 | 3.210 | 0.187 | 0.187 | XRD Nelson-Riley regression | EDS |
| Nb.24Mo.27Ta.25W.24 [44] | 3.214 | 3.229 | 3.215 | 0.467 | 0.031 | XRD Nelson-Riley regression | EDS |
| Nb.24Mo.18Ta.36W.22 [44] | 3.228 | 3.245 | 3.234 | 0.527 | 0.186 | XRD Nelson-Riley regression | EDS |
| Nb.216Mo.230Ta.281W.273 [45] | 3.2034 | 3.2303 | 3.2177 | 0.840 | 0.446 | XRD | EDS |
| Ti2ZrHfV0.5Mo0.2 [46] | 3.4584 | 3.3805 | 3.3845 | 2.252 | 2.137 | XRD | MNR |
| Ti.262Nb.255Ta.121Zr.242Al.120 [47] | 3.355 | 3.363 | 3.360 | 0.238 | 0.149 | XRD | EDS |
| V.098Co.301Cr.095Fe.455Ni.051 [48] | 3.582 | 3.610 | 3.604 | 0.782 | 0.614 | XRD | WCA |
| V.333Cr.309Fe.308Ta.025W.025 [49] | 2.935 | 2.948 | 2.930 | 0.443 | 0.170 | XRD Bragg´s Law | EDS |
| Co.286Al.071Fe.286Ni.286Mn.071 [50] | 3.6084 | 3.6061 | 3.5923 | 0.064 | 0.446 | XRD | EDS |
| CoCrFeNiPd [51] | 3.6473 | 3.6455 | 3.6658 | 0.049 | 0.507 | XRD | MNR |
| CoCrFeNiPd [52] | 3.6803 | 3.6455 | 3.6658 | 0.946 | 0.394 | XRD full spectrum fitting | MNR |
| Co.244Cr.244Fe.244Ni.244Al.024 [53] | 3.58 | 3.59 | 3.59 | 0.279 | 0.279 | XRD | EDS |
| Co.314Al.029Fe.3181i.307Mn.032 [54] | 3.5862 | 3.5798 | 3.5796 | 0.178 | 0.184 | XRD | EDS |
| Co.290Al.067Fe.288Ni.268Mn.087 [54] | 3.600 | 3.606 | 3.593 | 0.164 | 0.197 | XRD | EDS |
| Co.2Al.1Fe.3Ni.4 [55] | 3.5936 | 3.6132 | 3.5936 | 0.545 | 0.000 | XRD | MNR |
| NbMoTaVTi [56] | 3.1945 | 3.2153 | 3.2055 | 0.651 | 0.344 | XRD Rietveld | MNR |
| Nb.199Hf.198Ta.175V.212Ti.217 [57] | 3.279 | 3.295 | 3.290 | 0.488 | 0.336 | XRD | EDS |
| Nb.304Mo.037Ta.0511Zr.29Ti.319 [58] | 3.285 | 3.381 | 3.368 | 2.922 | 2.527 | XRD at higher angle peaks | EDS |
| Nb.255Mo.207Ta.19Zr.131Ti.217 [58] | 3.24 | 3.31 | 3.28 | 2.160 | 1.235 | XRD at higher angle peaks | EDS |
| Nb.090Mo.0247Ta.032Zr.566Ti.289 [58] | | 3.36 | 3.34 | 1.818 | 1.212 | TEM | EDS |
| Nb.275Mo.0947Ta.0911Zr.256Ti.284 [58] | 3.40 | 3.47 | 3.46 | 2.059 | 1.765 | TEM | EDS |
| NbZrHfVTi [59] | 3.3663 | 3.3613 | 3.3582 | 0.148 | 0.241 | Rietveld fullprof | MNR |
| Ir.191Pt.195Pd.207Rh.199Ru.208 [60] | 3.856 | 3.849 | 3.851 | 0.182 | 0.130 | XRD Rietveld | XRF |
| Co.237Cr.232Fe.245Ni.209Mn.077 [61] | 3.58 | 3.59 | 3.59 | 0.279 | 0.279 | XRD | EDS |
| Nb.238V.245Al.266Ti.251 [62] | 3.18 | 3.07 | 3.1 | 3.459 | 2.516 | XRD | MNR |
| Co.269Cr.250Fe.249Ni.184Mo.048 [63] | 3.585 | 3.603 | 3.606 | 0.502 | 0.586 | XRD (111) peak | EDS |
| Co.247Cr.245Fe.239Ni.246Mo.024 [64] | 3.604 | 3.589 | 3.594 | 0.416 | 0.278 | XRD Rietveld | EDS |
| Co.238Cr.2380Fe.238Ni.238Mo.048 [65] | 3.595 | 3.599 | 3.602 | 0.111 | 0.195 | Neutron diffraction | MNR |
| Co.204Cr.197Fe.299Ni.299 [66, 67] | 3.5759 | 3.5764 | 3.5782 | 0.014 | 0.064 | XRD (311) peak | EDS |
| Co.305Cr.208Fe.193Ni.294 [66, 67] | 3.5695 | 3.5704 | 3.5721 | 0.025 | 0.073 | XRD (311) peak | EDS |
| Co.296Cr.213Fe.303Ni.188 [66, 67] | 3.5741 | 3.5801 | 3.5822 | 0.168 | 0.227 | XRD (311) peak | EDS |
| Co.265Cr.205Fe.271Ni.260 [66, 67] | 3.573 | 3.576 | 3.578 | 0.084 | 0.140 | XRD (311) | EDS |
| | | | | | | peak | |
| Co.220Cr.244Fe.230Ni.307 [66, 67] | 3.5737 | 3.5758 | 3.577 | 0.059 | 0.092 | XRD (311) peak | EDS |
| Co.229Cr.236Fe.325Ni.210 [66, 67] | 3.5795 | 3.5834 | 3.5843 | 0.109 | 0.134 | XRD (311) peak | EDS |
| Co.315Cr.245Fe.245Ni.220 [66, 67] | 3.5704 | 3.6079 | 3.609 | 1.050 | 1.081 | XRD (311) peak | EDS |
| Co.216Cr.227Fe.276Ni.281 [66, 67] | 3.5752 | 3.5779 | 3.5792 | 0.076 | 0.112 | XRD (311) peak | EDS |
| Co.277Cr.233Fe.233Ni.277 [66, 67] | 3.572 | 3.5998 | 3.6013 | 0.778 | 0.820 | XRD (311) peak | EDS |
| Co.273Cr.229Fe.284Ni.214 [66, 67] | 3.5751 | 3.5801 | 3.5815 | 0.140 | 0.179 | XRD (311) peak | EDS |
| Co.188Cr.214Fe.198Ni.400 [66, 67] | 3.5708 | 3.5692 | 3.5711 | 0.045 | 0.008 | XRD (311) peak | EDS |
| Co.201Cr.193Fe.400Ni.205 [66, 67] | 3.5803 | 3.5847 | 3.5864 | 0.123 | 0.170 | XRD (311) peak | EDS |
| Co.248Cr.244Fe.251Ni.257 [66, 67] | 3.5767 | 3.5783 | 3.5793 | 0.045 | 0.073 | XRD (311) peak | EDS |
| Co.386Cr.196Fe.211Ni.208 [66, 67] | 3.568 | 3.5723 | 3.5744 | 0.120 | 0.179 | XRD (311) peak | EDS |
| | | | | | | | |

WCA = Wet chemical analysis.
MNR = Method used in analysis is not reported.
XRD* = X ray diffraction, plot of $h^2+k^2+l^2+$ vs $4Sen^2\theta/\lambda^2$.
WPPM = Whole powder pattern modelling.
EDS = Energy dispersive spectroscopy.
XRD = X ray diffraction.
TEM = Transmission electron microscopy.
XRF = X ray fluorescence.
ICP-OES = Inductively coupled plasma -optical emission spectrometry.
EPMA = Electron probe microanalysis

Table 3. Volume size factors, $\Omega_{sf}$ %, for systems not previously reported.

| Solvent | Solute | $\Omega_{sf}$ | | Solvent | Solute | $\Omega_{sf}$ |
|---|---|---|---|---|---|---|
| Al | Fe [68] | -47.65 | | Os | Ir [88] | 1.74 |
| | Mo [69] | -34.05 | | | Pd [*] | 4.77 |
| | Nb [70] | 4.48 | | | Pt [89] | 7.80 |
| | Ni [71] | -22.60 | | | Rh [90] | -1.63 |
| | | | | | | |
| Co | V [72] | 4.54 | | Pt | Os [92] | -7.12 |
| | Zn [73] | 43.57 | | | | |
| | | | | Re | Co [92] | -26.46 |
| Cr | Cu [74] | 3.33 | | | Fe [93] | -23.94 |
| | Ti [75] | 46.17 | | | Ir [94] | -5.61 |
| | Zn [*] | 26.50 | | | Rh [95] | -6.70 |
| | | | | | | |
| Fe | Ta [76] | 38.06 | | Rh | Os [96] | 1.58 |
| | | | | | Re [97] | 4.17 |
| Hf | Nb [77] | -19.95 | | | Ru [98] | 0.23 |
| | Ta [78] | -19.63 | | | | |
| | Ti [*] | -21.24 | | Ru | Co [99] | -18.28 |
| | V [*] | -40.81 | | | Fe [100] | -11.24 |
| | Zr [*] | 4.31 | | | | |
| | | | | Ta | Al [101] | -8.73 |
| Ir | Re [79] | 0.19 | | | Fe [102] | -38.9 |
| | Os [80] | -0.11 | | | Hf [103] | 21.93 |
| | | | | | | |
| Mo | Al [81] | -9.16 | | Ti | Nb [104] | 1.49 |
| | Co [83] | -30.11 | | | W [105] | -17.11 |
| | Fe [83 | -26.17 | | | | |
| | Ni [84] | -31.67 | | V | Hf [106] | 64.64 |
| | Zr [85] | 23.11 | | | Ni [107] | -27.83 |
| | | | | | Ti [*] | 26.22 |
| Nb | Al [86] | -9.42 | | | | |
| | Cr [87] | -28.97 | | Zn | Cr [*] | -36.15 |
| | | | | | | |
| | | | | Zr | Ta [*] | -25.02 |

* Interpolated from the least square linear equation of $\Omega_{sf}$ vs. solute atomic volume, for reported $\Omega_{sf}$ values.

The first term in (10) is the equivalent of Vegard´s law applied to HEA´s. It is the unit cell volume produced by the average atomic volumes of the different elements in the alloy. The second term in (10) would represent the volume changes produced during alloy formation. On average, for the alloys analyzed in this work, it represents 1.27% of the reported experimental unit cell volumes. For ten alloys, it was between 2.51%, and 7.77%, and for 23 alloys it was higher than 1% of the average unit cell volume. So, volume changes produced during formation of HEA´s must be considered when accurate values of cell volume or lattice parameters are required.

## 4.- Conclusions.

An equation has been derived to predict unit cell volume of high entropy alloys by two different methods. In the first method, a HEA is considered as a mixture of solid solutions. Based on an equation previously reported to calculate unit cell volume of intermetallic compounds, an expression to calculate unit cell volume of HEA´s was obtained. In the second method, a HEA is modeled as a mixture of solute element atoms surrounded by different environments. Using the effective volume of solute atoms in these environments, the unit cell volume of a HEA was derived. Both treatments led to the same equation. For cubic HEA´s lattice parameters were calculated. The predicted lattice parameters were compared with those reported for 68 HEA´s. Lattice parameters were also calculated using the equivalent of Vegard´s law for these alloys. Average errors were 0.52%, and 0.42% when Vegard´s law, and the equation derived in this work were used, respectively. Although these average errors seem acceptable, they were as high as 3.46%, and 2.52% when Vegard´s law, and the equation proposed are applied, respectively. The roles on error, of method to determine chemical composition, method to measure lattice parameter, and type of elements present on the alloy were examined.

**5.- References.**